# Multiple-scale analysis for resonance reflection by a one-dimensional rectangular barrier in the Gross-Pitaevskii problem

H. A. Ishkhanyan and V. P. Krainov

*Moscow Institute of Physics and Technology, 141700 Dolgoprudny, Moscow Region, Russia*

We consider a quantum above-barrier reflection of a Bose-Einstein condensate by a one-dimensional rectangular potential barrier, or by a potential well, for nonlinear Schrödinger equation (Gross-Pitaevskii equation) with a small nonlinearity. The most interesting case is realized in resonances when the reflection coefficient is equal to zero for the linear Schrödinger equation. Then the reflection is determined only by small nonlinear term in the Gross-Pitaevskii equation. A simple analytic expression has been obtained for the reflection coefficient produced only by the nonlinearity. An analytical condition is found when common action of potential barrier and nonlinearity produces a zero reflection coefficient. The reflection coefficient is derived analytically in the vicinity of resonances which are shifted by nonlinearity.

For studying quantum transmission and reflection, it is the most direct way to find exact solutions of the Schrödinger equation that dominates the dynamics of systems. However, only in a few cases with the simplest potentials, like rectangular well, the Schrödinger equation can be solved exactly. In most circumstances, exact solutions are difficult to obtain due to not only the effect of external field on particles, but also the interaction of particles. The most direct generalization of single-particle case is a tunneling of mean field through a barrier in the Gross-Pitaevskii, or nonlinear Schrödinger equation [1,2]. We emphasize that this is a nonlinear tunneling problem in the mean-field approximation. There have been various theoretical studies. From the theoretical point of view, the main complication in description of a quasistationary scattering process of particles obviously comes from the presence of atom-atom interaction. In leading order, the effect of this interaction is included in a nonlinear term in the Schrödinger-like Gross-Pitaevskii equation for wave function, using the Hartree self-consistent approximation with zero range interaction potential between atoms. The dynamics of solutions of this equation is very complex and rich. The phenomena of instabilities, focusing, and blowup are all concepts related to the nonlinear nature of the systems. Low velocity quantum reflection of Bose-Einstein condensates (BEC) of ultracold $^{23}$Na atoms from the attractive Casimir-Polder potential of silicon surface was observed experimentally in Refs. [3,4]. The measured reflection probability is in agreement with the theoretical model. Direct observation of tunneling and nonlinear self-trapping in a single bosonic Josephson junction was observed in Refs. [5,6]. Their results verify the predicted nonlinear generalization of tunneling oscillations in superconducting and superfluid Josephson junctions for two weakly linked Bose-Einstein condensates in a double-well potential. One of the first papers addressing nonlinear resonant tunneling of a BEC has been written by Paul *et al.* [7]. The most promising results for tunneling experiments are obtained using atom-chip-based waveguide interferometry with cold atom and Bose-Einstein condensate (see review [8]). A survey of nonlinear effects in BECs is given by Carretero-Gonzalez [9].

A convenient theoretical approach is based on the one-dimensional Gross-Pitaevskii equation

$$i\hbar \frac{\partial \psi(x,t)}{\partial t} = \left( -\frac{\hbar^2}{2m} \frac{\partial^2}{\partial x^2} + V(x) + g|\psi(x,t)|^2 \right) \psi(x,t),$$

which describes the dynamics in the mean-field approximation at low temperatures. Another important application is the propagation of electromagnetic waves in nonlinear media. The ansatz $\psi(x,t) = \exp(-i\mu t/\hbar)\psi(x)$ reduces the Gross-Pitaevskii equation to the corresponding time-independent (stationary) nonlinear Schrödinger equation with the chemical potential $\mu$

$$\left( -\frac{\hbar^2}{2m} \frac{d^2}{dx^2} + V(x) + g|\psi(x)|^2 \right) \psi(x) = \mu \psi(x). \quad (1)$$

In order to consider solutions in a finite trap, the rectangular well-potential

$$V(x) = \begin{cases} -V, & 0 < x < d; \\ 0, & x < 0, \ x > d \end{cases} \quad (2)$$

is studied. This potential gives analytic solutions, unlike the harmonic traps. It also provides a model for a finite depth trap. The treatment of transport within this mean-field theory reveals interesting phenomena which arise from the nonlinearity of the equation. We consider also the case $V<0$ which corresponds to the potential barrier.

As already shown in Refs. [10,11] a solution of Eq. (1) with the potential (2) is given in terms of the Jacobi elliptic functions $dn(x)$. The Gross-Pitaevskii equation for a one-dimensional finite rectangular well potential was studied in Ref. [12]. in terms of ingoing and outgoing waves. The transmission coefficient varies periodically as a function of the chemical potential $\mu$. Thus, in resonances there is total transparency of the potential barrier, or of the potential well. The resonance line shape was investigated in recent papers [13,14].

A particle moving through a classically allowed region can be reflected by the potential without reaching a classical turning point. Above-barrier reflection occurs also when $V<0$ and the chemical potential $\mu$ is larger than $|V|$. In the linear problem [$g=0$ in Eq. (1)] with the potential (2), the reflection coefficient $R$ is determined by expression ([15,16])

$$R = \frac{(k_1^2 - k_2^2)^2 \sin^2(k_2 d)}{(k_1^2 - k_2^2)^2 \sin^2(k_2 d) + 4(k_1 k_2)^2}. \quad (3)$$

Here, $k_1 = \sqrt{2\mu}$ and $k_2 = \sqrt{2(\mu+V)}$. $\mu > |V|$ at the above-barrier transmission and reflection. We use everywhere the system of units $\hbar = m = 1$. It is seen in linear problem that $R = 0$ when $k_2 d = n\pi$ and $n = 1, 2, 3 \ldots$. The resonances defined by the condition $R = 0$ are shifted by the nonlinearity as discussed in Refs. [7,12].

We use the multiple-scale analysis for derivation of the resonant reflection coefficient in order to avoid secular terms [17]. This approach was applied for consideration of Bose-Einstein solitons in highly asymmetric traps [18]. Quantum reflection of the incident soliton by an attractive sech-squared-shape potential $V(x) = -V \text{sech}^2(x/x_0)$ (Rosen-Morse potential) was analyzed in [19]. It was shown in [20] that the well-known absolute transmission of the nonlinear system can occur also in the Rosen-Morse potential. It is possible that absolute transmission of the nonlinear system can occur also in many other potentials (see discussion in Ref. [21]).

The nonlinearity is assumed to be small, i.e., $g \ll \mu$. Then for the reflection coefficient for rectangular well, or barrier it is possible to find a simple analytic expression using multi-scale approach in the vicinity of resonances. This is just the goal of our work.

We assume that motion of Bose-Einstein condensate takes place in positive direction of the axis $X$. Then at $x < 0$ there are both incident and reflecting wave while at $x > d$ there is only the transmitting one. The most simple is the consideration of the region $x > d$. We introduce the dimensionless nonlinearity parameter $a = g/\mu$. Repulsive nonlinearity occurs at $a > 0$ while attractive one occurs at $a < 0$. The solution of Eq. (1) is of the form

$$\psi(x) = E \exp(i\kappa x), \quad (4)$$

where

$$\kappa = k_1 \sqrt{1 - a|E|^2}. \quad (5)$$

When $k_2 d = n\pi$ and the reflection coefficient $R = 0$, in linear problem the constant $E$ is equal to

$$E = E_0 = (-1)^n \exp(ik_1 d). \quad (6)$$

In nonlinear problem with small nonlinearity, we restrict ourselves to the first-order terms in $a \ll 1$. Then $E = E_0(1+e)$ where $e \ll 1$ is the complex quantity which will be determined by the matching conditions. Hence, it follows from Eq. (5) that $\kappa \approx k_1(1 - a/2)$.

Now we consider the region $x < 0$. Let us change the independent variable as $y = k_1 x$. Then the Gross-Pitaevskii equation, Eq. (1), at $x < 0$ is of the form

$$\frac{d^2 \psi}{dy^2} + \psi = a|\psi|^2 \psi. \quad (7)$$

The multiple-scale analysis is one of the versions of the perturbation theory [17]. Let us introduce the new independent variables $y_1 = y$, $y_2 = ay$, $y_3 = a^2 y \ldots$ Then

$$\frac{d\psi}{dy} = \frac{\partial \psi}{\partial y_1} + a \frac{\partial \psi}{\partial y_2} + a^2 \frac{\partial \psi}{\partial y_3} + \ldots;$$

$$\frac{d^2 \psi}{dy^2} = \frac{\partial^2 \psi}{\partial y_1^2} + 2a \frac{\partial^2 \psi}{\partial y_1 \partial y_2} + \ldots$$

Let us expand further wave function into a series of the small parameter $a \ll 1$: $\psi = \psi_0 + a\psi_1 + a^2 \psi_2 + \ldots$ The equation for the zero approximation wave function $\psi_0$ is of the form $\partial^2 \psi_0 / \partial y_1^2 + \psi_0 = 0$ and its general solution is (we restrict ourselves by two independent variables $y_1$ and $y_2$)

$$\psi_0(y_1, y_2) = c(y_2) \exp(iy_1) + b(y_2) \exp(-iy_1). \quad (8)$$

The inhomogeneous differential equation for the first approximation wave function $\psi_1$ is

$$\frac{\partial^2 \psi_1}{\partial y_1^2} + \psi_1 = |\psi_0|^2 \psi_0 - 2 \frac{\partial^2 \psi_0}{\partial y_1 \partial y_2}. \quad (9)$$

The idea of a multiple-scale analysis is to remove secular terms in Eq. (9). Secular terms on the right side of Eq. (9) are solutions of the corresponding homogeneous equation. Substituting Eq. (8) into Eq. (9), one obtains

$$\frac{\partial^2 \psi_1}{\partial y_1^2} + \psi_1 = \{|c|^2 + |b|^2 + cb^* \exp(2iy_1) + c^* b \exp(-2iy_1)\}$$

$$\times [c \exp(iy_1) + b \exp(-iy_1)]$$

$$- 2i \frac{dc}{dy_2} \exp(iy_1) + 2i \frac{db}{dy_2} \exp(-iy_1). \quad (10)$$

Presenting the complex variables $c(y_2)$ and $b(y_2)$ in the standard form $c = C \exp(i\phi)$ and $b = B \exp(i\varphi)$ where $C, B$ and $\phi, \varphi$ are real functions of $y_2$, we require in accordance with multiple-scale analysis that the right side of Eq. (10) should not contain the secular terms of the type $\exp(iy_1)$ and $\exp(-iy_1)$, which are simultaneously the solutions of the left side of Eq. (10). When the nonlinearity is small, the coefficient $C$ before $\exp(iy_1)$ can be chosen to be equal to 1 analogously to the corresponding choice in the linear problem for continuum wave functions. Then it follows from Eq. (10) equations for $B$, $\phi$, and $\varphi$:

$$1 + 2B^2 = -2 \frac{d\phi}{dy_2}; \quad B(B^2 + 2) = -2i\left(\frac{dB}{dy_2} + iB \frac{d\varphi}{dy_2}\right).$$

The simple solutions of these equations are of the form $B = \text{const} \sim a$, and $\phi = -(B^2 + 1/2)y_2 \approx -ak_1 x/2$; $\varphi = (1 + B^2/2)y_2 + \pi/2 \approx ak_1 x + \pi/2$. Here, we neglect here the terms which are on the order of $a^2$.

Thus, the solution, Eq. (8) at $x < 0$ taking into account the zero approximation terms can be written of the form $\psi_0(x) = \exp(i\nu_1 x) + iB \exp(-i\nu_2 x)$, where the notations are introduced $\nu_1 = k_1(1 - a/2)$; $\nu_2 = k_1(1 - a)$.

We retain in wave function only terms which are of the zero and first order in the small parameter $a$. It follows from Eq. (10) that $\psi_1 \sim B \sim a$. The function $\psi_1$ contains third harmonics $\exp(3ik_1 x)$ and $\exp(-3ik_1 x)$. Since $a\psi_1 \sim a^2$ we neglect the function $a\psi_1$ in comparison to $\psi_0$. Hence, when $x < 0$,

$$\psi(x) = \exp(i\nu_1 x) + iB \exp(-i\nu_2 x). \quad (11)$$

The quantity $B \ll 1$ will be determined from the matching conditions. Then the quantity $R = B^2$ is the reflection coefficient.

It should be noted that such an approach is valid only when the parameter of nonlinearity $a$ is small. When $|a|>1$, terms $a_3 \exp(3ik_1x)$, $a_5 \exp(5ik_1x),\ldots$ $b_3 \exp(-3ik_1x)$, $b_5 \exp(-5ik_1x),\ldots$ should be added to Eq. (11). Therefore, the averaged reflecting current will be of the form $j_{\text{ref}} = \nu_2|b|^2 + 3k_1|b_3|^2 + 5k_1|b_5|^2 + \ldots$ and the quantity $\nu_2$ should take into account the terms of higher orders in the nonlinearity parameter $a$. The incident current $j_{\text{inc}}$ is also modified by analogy. Then the reflection coefficient can be determined as $R = j_{\text{ref}}/j_{\text{inc}}$. We postulated here the ansatz that when we expand an arbitrary function into Fourier series, then exponents with positive wave numbers $k$, $3k$, $5k\ldots$ correspond to incident wave function, while exponents with negative wave numbers $-k$, $-3k$, and $-5k\ldots$ correspond to reflected wave function. Indeed, in contrast to the case of the linear Schrödinger equation the transmission coefficient cannot be computed by simply decomposing the wave function into an incident and a reflected part because the superposition principle of quantum mechanics is not valid in the presence of the nonlinear term. However, such a decomposition is possible in the limit of a small nonlinearity, or small back reflections. It should be noted that the ansatz $\psi(x) = A_1 \exp(ik_1x) + B_1 \exp(-ik_1x)$, $x<0$ suggested in Ref. [12]. [see Eq. (4) of this reference] is correct since they consider a model where $g=0$ outside of the square-well. Another definition of the transmission coefficient was suggested by Paul et al. [22]: the transmission coefficient is evaluated by the ratio of the current in the presence of the potential $V(x)$ (i.e., the transmitted current) to the current obtained in the absence of $V(x)$ (the incident current that is emitted by the source). Authors of the paper [23] follow an ansatz closely related to usual experimental setups, and choose to work with an incident and a reflected beam, which can be approximated by plane waves. This corresponds to a regime where suggested by Leboeuf et al. a semiclassical Schrödinger-like equation for the amplitude of the wave function can be linearized in the far upstream region. They solve numerically the exact nonlinear equation and use the linearization procedure only to define the transmission coefficient. In particular, they consider an abrupt steplike constriction. Another approach for description of nonlinear tunneling suggested by Deckel et al. [24] is to postulate that initial Gaussian wave packet whose peak is situated at the center of the potential well at the time $t=0$ evolves dynamically according to the time-dependent Gross-Pitaevskii equation.

Now we consider the region $0<x<d$. Omitting the details of derivations by the multiple-scale analysis, which were described above in detail for the case $x<0$, one obtains

$$\psi(x) = \left[\frac{k_2+k_1}{2k_2} + c\right]\exp(i\nu_3 x) + \left[\frac{k_2-k_1}{2k_2} + h\right]\exp(-i\nu_4 x)$$
$$- \frac{a}{64k_2^3}(k_2^2 - k_1^2)[(k_2+k_1)\exp(3ik_2x)$$
$$+ (k_2-k_1)\exp(-3ik_2x)]. \quad (12)$$

Here, the notations are introduced

$$\nu_3 = k_2\left[1 - a\frac{3k_1^2 + 3k_2^2 - 2k_1k_2}{8k_2^2}\right];$$

$$\nu_4 = k_2\left[1 - a\frac{3k_1^2 + 3k_2^2 + 2k_1k_2}{8k_2^2}\right].$$

The constants $c \ll 1$ and $h \ll 1$ are determined from the matching conditions. When $c=h=a=0$ Eq. (12) is reduced to the well-known corresponding wave function in the linear problem for region $0<x<d$. Unlike the region $x<0$, here we cannot neglect third harmonics since they are linear functions of the small nonlinearity parameter $a$.

Now we match the wave functions, Eqs. (11) and (12) at $x=0$. Terms of the first order in $a$ give the relation

$$h + c - iB = \frac{a}{32k_2^2}(k_2^2 - k_1^2). \quad (13)$$

By analogy, matching of the first derivatives of these functions at $x=0$ results in the relation

$$\frac{k_2}{k_1}(h - c) - iB = \frac{9a}{32k_2^2}(k_2^2 - k_1^2). \quad (14)$$

Matching of the wave functions, Eq. (12) and (4), at $x=d$ is of the form

$$h + c - e = \frac{a}{32k_2^2}(1 - 12ik_1d)(k_2^2 - k_1^2). \quad (15)$$

Finally, matching of the first derivatives of these functions at $x=d$ is of the form

$$\frac{k_2}{k_1}(h - c) + e = \frac{3a}{32k_2^2}\left(3 - 4i\frac{k_2}{k_1}k_2d\right)(k_2^2 - k_1^2). \quad (16)$$

Thus, one obtains the system of four algebraic linear equations, Eqs. (13)–(16) for the quantities $B$, $c$, $h$, and $e$. The solution for $B$ is of the form

$$B = \frac{3ad}{16k_1k_2}(k_1+k_2)(k_1^2 - k_2^2). \quad (17)$$

The reflection coefficient is of a simple analytic form (taking into account that $k_2d = \pi n$ in resonance)

$$R = B^2 = \left(\frac{3gd^2(1+s)(1-s^2)}{8\pi n}\right)^2. \quad (18)$$

Here,

$$s = \frac{\pi n}{d\sqrt{2\mu}}. \quad (19)$$

When $V=0$ then $s=1$ and, of course, $R=0$. It should be noted that if $V>0$ (potential well) then the value of the integer $n$ should satisfy the inequality $n > \sqrt{2Vd}/\pi$. Then $\mu > 0$ and $s > \sqrt{V/\mu}$.

When $k_2d = n\pi + \delta$, and $\delta \neq 0$, but $\delta \ll 1$, the reflection amplitude $B_0$ in the linear problem ($g=0$) is nonzero. It is of the form

$$B_0 = (-1)^n \frac{Vd}{k_1n\pi}\delta \ll 1. \quad (20)$$

It follows from Eqs. (17) and (20) that total reflection amplitude in this case is

$$B_0 + B = \frac{Vd}{k_1n\pi}\left[(-1)^n\delta - a\frac{3(k_1d + \pi n)}{8}\right]. \quad (21)$$

Then the reflection coefficient is

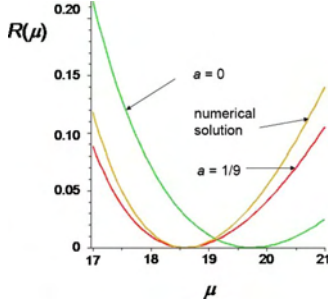

FIG. 1. (Color online) The dependence of the dimensionless reflection coefficient $R(\mu)$ on the chemical potential $\mu$ (in units $\hbar^2/md^2$) for the case $n=1$, and $V=-3\pi^2/2$. The nonlinear parameter is $a=1/9$, or $a=0$. In linear case ($a=0$) the quantity $R=0$ when $\mu=2\pi^2$. In nonlinear case ($a=1/9$) the quantity $R=0$ when $\mu=241\pi^2/128<2\pi^2$. The result of numerical solving of Gross-Pitaevskii equation for the same values of parameters is also shown.

$$R = (B_0 + B)^2 = \left(\frac{Vd}{k_1 n\pi}\right)^2\left[(-1)^n\delta - a\frac{3(k_1d+\pi n)}{8}\right]^2.$$

Let us consider the example: $n=d=1$, $a=1/9$, and $V=-3\pi^2/2$ (potential rectangular barrier). The usual units for $V$ and $\mu$ are $\hbar^2/md^2$. One obtains $\mu \approx 2\pi^2 > |V|$ so that

$$R(\mu) = \frac{9}{16}\left[\sqrt{2\mu - 3\pi^2} - \frac{7\pi}{8}\right]^2. \quad (22)$$

This dependence is shown in Fig. 1. It is seen that when $a>0$ the curve is shifted slightly to the left in comparison to the linear case, i.e., when $a=0$. The next resonance ($n=2$) is shifted to the right and so on.

In particular, $R=0$ when $\mu=241\pi^2/128<2\pi^2$. On Fig. 1 also the function $R_0(\mu)$ is shown for $a=0$ (linear case), i.e.,

$$R_0(\mu) = \frac{9}{16}[\sqrt{2\mu - 3\pi^2} - \pi]^2. \quad (23)$$

Dotted line in Fig. 1 is the result of numerical solution of Eq. (1) for the same values of parameters.

According to Eq. (21), the total transparency of barrier transmission ($R=0$, $T=1-R=1$) is realized under the condition (see also [12])

$$\delta = (-1)^n a\frac{3(k_1d + n\pi)}{8}. \quad (24)$$

Some comments can be made concerning the transmission amplitude $1+e$ in the case of exact resonance ($\delta=0$). It follows from Eqs. (13)–(16) that

$$e = \frac{3iad}{16k_1k_2}(k_2-k_1)(k_1^2-k_2^2). \quad (25)$$

Then formally the transmission coefficient $T=1+|e|^2>1$. This contradiction can be eliminated by multiplying the wave function in all regions by the factor $1/\sqrt{1+|e|^2}$. Since we restrict ourselves to terms which are linear in the nonlinearity parameter $a$, this procedure is correct, and the additional terms are proportional to $a^2$. In order to obtain correct transmission coefficient, we should further multiply the wave function by the factor $1/\sqrt{1+B^2}$. Then the reflection coefficient $R=B^2\ll 1$ will not change in the linear approximation, but the transmission coefficient takes the form $T=1-B^2$ as it should be.

In conclusion, the scattering of BEC by a finite rectangular well potential has been discussed in terms of stationary states of the Gross-Pitaevskii equation. Neglecting the mean-field interaction outside the potential, ingoing and outgoing waves together with reflection and transmission probabilities can be defined within the approximation of a weak nonlinear parameter using the multiple-scale analysis. The vicinity of resonances has been investigated where the role of the weak nonlinearity is significant. A simple analytical expression for the reflection coefficient in the case when reflection is absent in the linear problem as well as the reflection coefficient in the vicinity of resonances of the linear problem has been obtained. Positions of resonances have been found where the reflection coefficient is zero in the presence of both nonlinearity and some small detuning from resonance in the linear problem. Unfortunately, the asymmetry and the bending of resonances that appear when the chemical potential is sufficiently far from resonance point cannot be considered analytically.

Support from the RFBR, Project No. N 07.02.00080, is gratefully acknowledged. We thank Dr. L. Carr and Dr. H. J. Korsch for discussions and valuable comments.